\documentclass[journal=jacsat,manuscript=article]{achemso}

\usepackage[version=3]{mhchem} 

\usepackage{xcolor}

\author{J\"org Rottler}
\affiliation{Department of Physics and Astronomy and Stewart Blusson Quantum Matter Institute, University of British Columbia, Vancouver, British Columbia V6T 1Z1, Canada}
\email{jrottler@physics.ubc.ca}
\author{Christoph Ortner}
\affiliation{Department of Mathematics,  University of British Columbia, Vancouver, British Columbia V6T 1Z2, Canada}

\title[An \textsf{achemso} demo]
  {Analysis of local structure of mechanical and thermal rearrangements in glasses with the atomic cluster expansion}

\abbreviations{IR,NMR,UV}

\begin{document}

\begin{tocentry}

\includegraphics[width=8.2cm]{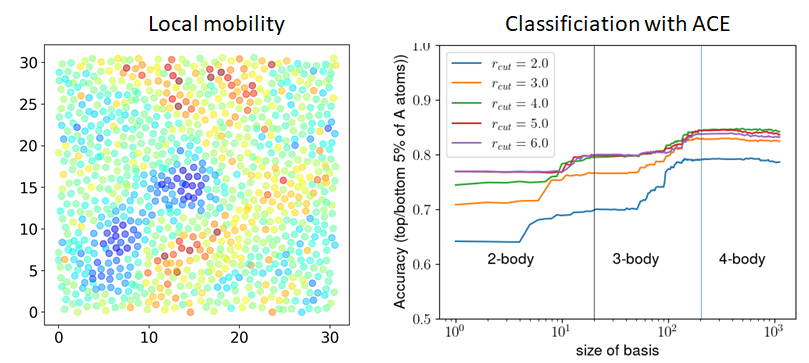}




\end{tocentry}

\begin{abstract}
We explore the structural signatures of excitations in amorphous materials with the atomic cluster expansion (ACE), a universal and complete linear basis of descriptors of the atomic environment. Body-orderd linear classifiers are constructed that distinguish between active and inactive particles in three different model glass formers, in which structural relaxation occurs either through spontaneous thermal activation or by simple shear. We find that in binary mixtures, maximum prediction accuracy is already achieved with very few two-body correlations, while a polymer glass requires both two- and three-body correlations. Trends are robust across both activation mechanisms.   

\end{abstract}

\section{Introduction}
When the temperature of a glass-forming material is lowered towards its glass transition, the molecular motion that is spatially uniform in the fluid state develops spatially and temporally varying features that are termed dynamic heterogeneity \cite{ediger2000spatially,berthier2011dynamical}. Regions of highly mobile particles begin to form next to regions of nearly stationary ones, and they coexist over a timescale of order the mean relaxation time of the material. A similar form of heterogeneous dynamics occurs in glassy solids under deformation. When subject to shear, plastic flow does not occur uniformly in the material, but global deformation is instead mediated via localized, short lived shear transformations that occur in an otherwise homogeneous background \cite{argon1979plastic}. 

A pertinent question that has attracted abiding interest is whether such localized dynamics can be correlated with a purely structural origin. The answer to this question, after extensive research efforts over the last 25 years, is a resounding yes. To this end, particle scale simulations have been particularly helpful. Correlations between structure and dynamics have for instance been established via scalar (bond) order parameters, Voronoi tessellation of particle configurations, or low frequency vibrational modes of the dynamical matrix if information from the interaction potential is also used \cite{richard2020predicting}. A productive approach has also been to construct a general basis of molecular descriptors of the local particle environment \cite{behler2007generalized,bartok2013}, and then use such features in classification or regression algorithms from supervised machine learning (ML) to predicting the probability of a local rearrangement; for a recent review, see ref.~\cite{jung2023roadmap}.

A universal and complete descriptor of an atomic environment is given by the atomic cluster expansion (ACE), introduced recently in the context of machine learning interatomic potentials for simulating complex materials with high fidelity \cite{drautz2019,dusson2022}. It provides a permutation, translation and rotation invariant description of the local environment of a particle that can be systematically improved by increasing the body order of the expansion.  By examining the convergence, it is thus possible to assess the importance of multibody correlations for predicting the target quantity of interest. 

ACE has so far not been employed for the analysis of glassy dynamics. Previous ML models have used other descriptor sets such as the symmetry functions introduced by Behler and Parrinello \cite{cubuk2015,schoenholz2016structural,wu2023machine}, angular descriptors inspired by bond-orientational order parameters \cite{boattini2021,alkemade2023improving} or the Smooth Overlap of Atomic Positions (SOAP) \cite{wang2020predicting,coslovich2022}. These capture two- and three-body correlations in the atomic environment, and can be shown to be either related or contained in the more general ACE formalism. While these models certainly succeed in providing good predictions for local molecular mobility with linear regression or support vector machines, unlike the ACE descriptors they are not systematic (i.e., complete). There have also been intensive efforts in further improving the structure-property predictions via deep learning \cite{bapst2020unveiling,fan2021predicting,shiba2023,jung2023predicting,oyama2023,pezzicoli2024,jung2024dynamic}, but such highly nonlinear methods come at the expense of reduced interpretability.  

A first goal of the present contribution consists in investigating the importance of higher order correlations in the link between structure and dynamics of glasses with the help of ACE. We also seek to assess the variation of the structure-dynamics correlations across different glass formers, and for different excitation mechanisms. We will do this by triggering relaxation processes in three different model systems in two different ways: (i) in the quiescent supercooled fluid at equilibrium, in which rearrangements are spontaneous and purely thermally activated and (ii) in the glass subject to simple quasistatic shear at zero temperature, in which rearrangements are purely mechanically activated and only happen when the system reaches an instability. While both scenarios have been studied separately in depth many times in the literature, we are not aware of a study that applies the same ML framework simultaneously to both dynamical processes in the same system, and using the same indicator of molecular mobility. In this way, we hope to illuminate any commonalities and differences in the glassy structure that underpins locally heterogeneous motion that is excited either mechanically or thermally. Throughout this paper, we restrict ourselves to linear models in the interest of simplicity and interpretability. 

\section{Methods}

\subsection{The atomic cluster expansion}

\begin{figure}[t]
  \includegraphics[width=10cm]{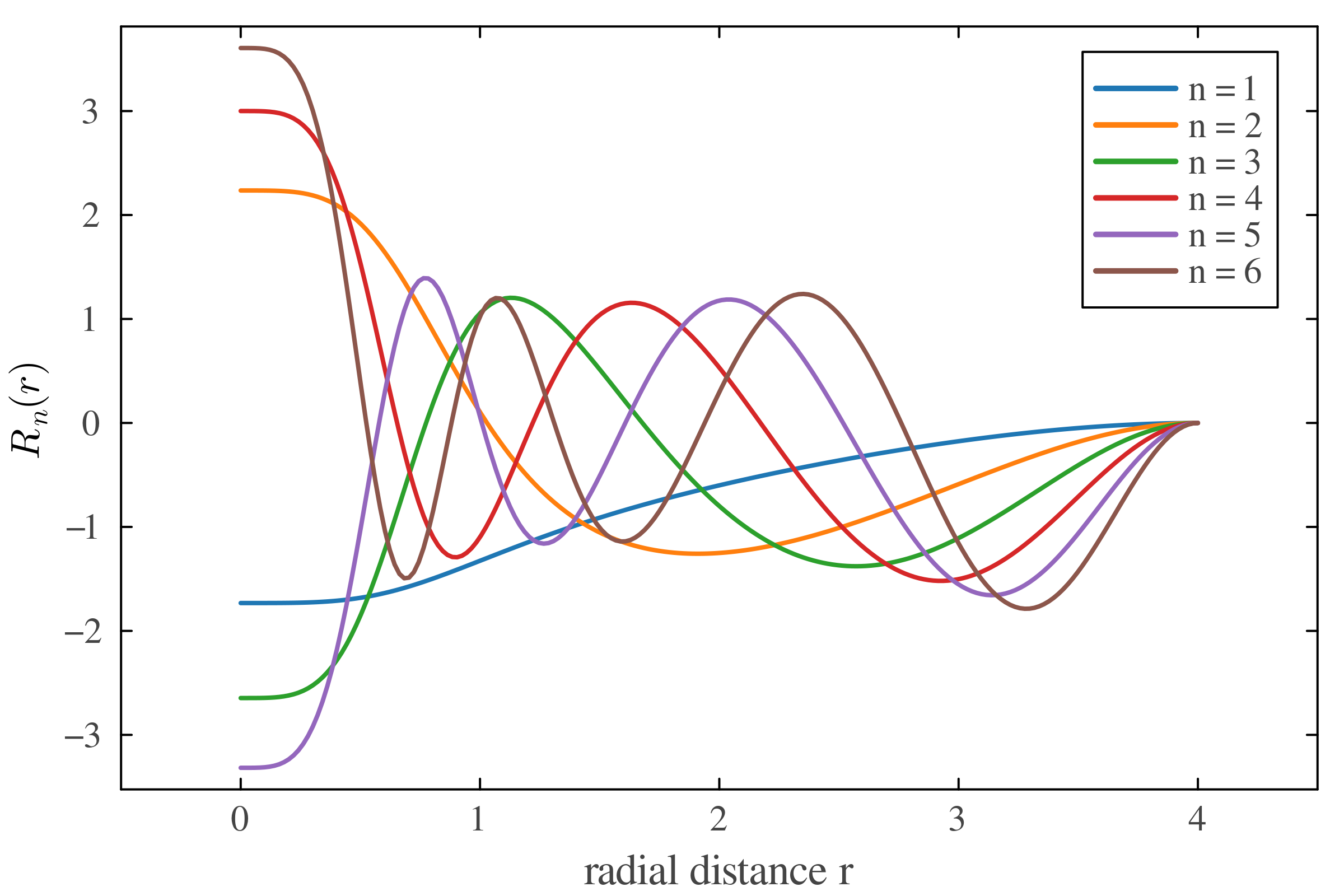} 
  \caption{The first 6 radial basis functions $R_n(r)$ used for the atomic cluster expansion features, for $r_{cut}=4$. The distance transform $x(r)$ results in increased resolution of the basis near $r = 1$. }
  \label{fgr:radials}
\end{figure}

The atomic cluster expansion (ACE) is described in detail in the original paper by Drautz \cite{drautz2019} and in subsequent work \cite{dusson2022,witt2023}. Here we provide only a synopsis of the most pertinent equations. We define the ``atomic density'' of atom $i$ as 
\begin{equation}
    \rho^{(i)}({\boldsymbol{r}, z}) := \sum_j^{N_c} \delta({\boldsymbol{r} } - {\boldsymbol{r} }_{ji}) \delta(z - z_j),
\end{equation}
where $\boldsymbol{r}_{ji}=\boldsymbol{r}_j-\boldsymbol{r}_i$, $z_j$ labels the species of particle $j$, and the sum runs over all neighboring particles $N_c$ within a cutoff radius $r_{cut}$.
Then we choose a one-particle basis 
\begin{equation}
    \phi_v({\boldsymbol{r}, z}) = \phi_{\zeta nlm}({\boldsymbol{r}, z}) = \delta_{\zeta z} R_n(r) Y_l^m(\hat{\boldsymbol{r} })
\end{equation}
where $R_n(r)$ denotes a radial basis built on a complete set of polynomials, and $Y_l^m(\hat{\boldsymbol{r}})$ the spherical harmonics (normally taken complex). While the latter are required to respect the relevant symmetries, considerable freedom exists in the choice of the radial basis. Here we use the construction suggested in ref.~\cite{witt2023}, 
\begin{equation}
    R_n(r)= E(x(r)) \cdot P_n(x(r)),
    \label{rn-eq}
\end{equation}
where $x(r)$ is a distance transform, $E(x) = x - x(r_{\rm cut})$ an envelope function and $P_n(x)$ denotes the Legendre polynomials rescaled to be orthonormal. The distance transform is chosen empirically as 
\[
x(r) = 
1 - \frac{y(r/r_{\rm cut}) - y(1)}{y(0) - y(1)} \cdot \bigg(1 - \frac{r^4}{r_{\rm cut}^4}\bigg), \qquad 
y(s) = \frac{1 + s^{3}}{1 + s^{3} + a s^4}.
\]
with $a$ chosen such that $|y'(r)|$ is maximal at $r = 1$, which maximizes resolution around the nearest-neighbour distance. The rationale for those choices is discussed in more details in \citet{witt2023}. 
%
%
%
Figure \ref{fgr:radials} shows the first 6 basis functions $R_n(r)$ constructed in this way.

We now project $\rho$ onto that basis, 
\begin{equation}
    A_{v}^{(i)} = \langle \phi_{v}, \rho^{(i)} \rangle = \sum_j \phi_{v}({\boldsymbol{r} }_{ji}, z_j).
\end{equation}
Next, we form the $N$-correlations of the density, 
\begin{equation}
    {\boldsymbol{A} }^{(i)}_{{\boldsymbol{v} }}
    = \Big\langle \otimes_{t=1}^N \phi_{v_t}, (\rho^{(i)})^{\otimes N} \Big\rangle = 
    \prod_{t = 1}^N A^{(i)}_{v_t}. 
\end{equation}
A vector of features of the atomic environment are then obtained by a symmetrization operation, 
\begin{equation}
    \boldsymbol{B}^{(i)}=\mathcal{C} \cdot \boldsymbol{A}^{(i)}
\end{equation}
where the coefficients $\mathcal{C}$ are chosen to select all possible linear combinations of $\boldsymbol{A}_{\boldsymbol{v}}^{(i)}$ features that are rotation and reflection invariant. 

Explicity, the first three groups of this expansion can be written as 
\begin{equation}    \boldsymbol{B}^{(i, 1)}_{\zeta, n}=A^{(i)}_{\zeta n00}=\sum_j^{N_c}R_n(r_{ji}) \delta_{\zeta z_j}
\end{equation}
\begin{equation}
    \boldsymbol{B}^{(i, 2)}_{\zeta_1 \zeta_2 n_1 n_2 l} = \sum_{m=-l}^l (-1)^m A^{(i)}_{\zeta_1 n_1lm}A^{(i)}_{\zeta_2 n_2l-m}
\end{equation}
\begin{equation}
    \boldsymbol{B}^{(i, 3)}_{\boldsymbol{\zeta n l}} = \sum_{m_1=-l_1}^{l_1}  \sum_{m_2=-l_2}^{l_2} \sum_{m_3=-l_3}^{l_3}      
    \left( {\begin{array}{ccc}
    l_1 & l_2 & l_3 \\
    m_1 & m_2 & m_3 \\
  \end{array} } \right)\times
    A^{(i)}_{\zeta_1 n_1l_1m_1}A^{(i)}_{\zeta_2 n_2l_2m_2}A^{(i)}_{\zeta_3 n_3l_3m_3},
\end{equation}
where the $()$ denote the Wigner 3j symbol. These features represent 2-, 3- and 4-body contributions in the cluster expansion. The feature vectors $B^{(i,2)}$ and $B^{(i,3)}$ are often also called powerspectrum and bispectrum (of the atomic density). Note that the 2-body terms can also be interpreted as a coarse-grained local density, with the polynomial basis $R_n(r)$ acting as a weighting function.
The features can now be used in a linear regression model that can be fitted against any target quantity $T_i$ defined at the individual particle level, 
\begin{equation}
    T_i=\boldsymbol{\Theta}\cdot\boldsymbol{B}^{(i)}.
\end{equation}
In the present study, we will be performing a binary classification of particles into active $(T_i=1)$ and inactive $(T_i=0)$ particles. In this case, logistic regression is more appropriate, where the probability of observing an active particle is given by the sigmoid function 
\begin{equation}
    \label{eq:classifier}
    P(T_i=1|{B}^{(i)},\boldsymbol{\Theta}) = \sigma\big( \boldsymbol{\Theta} \cdot \boldsymbol{B}^{(i)}\big) 
    := 
    \frac{1}{1 + e^{-\boldsymbol{\Theta}\cdot\boldsymbol{B}^{(i)}}}.
\end{equation}
Interpretability of this ACE model is given by rewriting it in terms of a self-interacting cluster expansion, 
\begin{equation}
    \label{eq:self-interacting}
    \begin{aligned}
     \boldsymbol{\Theta} \cdot \boldsymbol{B}^{(i)} 
    = 
     & \sum_{j_1} u_1\big(z_i; \boldsymbol{r}_{ij_1}, z_{j_1}\big)
        + \sum_{j_1, j_2} u_2\big(z_i; \boldsymbol{r}_{ij_1}, z_{j_1}, 
                        \boldsymbol{r}_{i j_2}, z_{j_2} \big) \\ 
    & + \dots +  \sum_{j_1,\dots,j_N} 
            u_N\big( z_i; \boldsymbol{r}_{i j_1}, z_{j_1}, \dots,
                        \boldsymbol{r}_{i j_N}, z_{j_N} \big) ,
    \end{aligned}
\end{equation}
where $N+1$ is the body-order of the model, each sums $\sum_{j_t}$ runs over all neighbours of a central atom $i$, and the functions $u_n, n = 1, \dots, N$ are implicitly given by the parameterization; see \citet{witt2023} for the details.  

A potential downside of \eqref{eq:self-interacting} for interpretability is the fact that the $(n+1)$-body terms are self-interacting; that is the summation includes spurious clusters with repeated atoms. For example, the 3-body and 4-body terms modify the 2-body, and the 4-body term modifies the 3-body. To avoid this effect, we employ a purification procedure that removes those self-interaction terms.\cite{dusson2022,witt2023, pureace2024} It can be shown\cite{pureace2024} that for the specific choice of radial basis that we employ a modification of the coupling operator $\mathcal{C}$ that leads to an alternative invariant feature vector for which the interpretation is the more common many-body expansion, 
\begin{equation}
    \label{eq:pureace}
    \begin{aligned}
     \boldsymbol{\Theta} \cdot \boldsymbol{B}^{(i)}
    = 
     & \sum_{j_1} v_1\big(z_i; \boldsymbol{r}_{ij_1}, z_{j_1}\big)
        + \sum_{j_1 < j_2} v_2\big(z_i; \boldsymbol{r}_{ij_1}, z_{j_1}, 
                        \boldsymbol{r}_{i j_2}, z_{j_2} \big) \\ 
    & + \dots +  \sum_{j_1 < \dots < j_N} 
            v_N\big( z_i; \boldsymbol{r}_{i j_1}, z_{j_1}, \dots,
                        \boldsymbol{r}_{i j_N}, z_{j_N} \big) ,
    \end{aligned}
\end{equation}
where the summation, $\sum_{j_1 < \dots < j_N}$, now only involved unique and physical $N$-clusters.
In our numerical tests we employ this {\em purified cluster expansion} model. 

An important advantage of the ACE model is that its computational cost scales linearly with the number of neighbors $N_c$ in the cutoff range irrespective of the order $N$ of the expansion, in contrast to the $N_c^N$-scaling of a naive implementation of the cluster expansion \eqref{eq:pureace}. The accuracy of the expansion can be tuned by increasing the cutoff radius, body-order, and polynomial basis. ACE descriptors are computed with a variant of the code presented in Witt et al.~\cite{witt2023}.

\subsection{Glass models and molecular dynamics simulation details}
In order to assess the strength of the structure-dynamics correlation between different glass formers, we consider three different model systems. Our first system is the FENE bead-spring homopolymer model, where $N=30,000$ particles interacting via the Lennard-Jones (LJ) potential are coupled together with nonlinear springs to form 1,500 linear chains of 20 momoners each \cite{kremer1990dynamics}. This monatomic glass is complemented by two binary mixtures representing simple metallic glasses, namely the Kob-Andersen (KA) system with the composition 80-20 \cite{kob1995testing} and the Wahnstrom mixture with the composition 50-50 \cite{wahnstrom1991molecular}. The interaction potential is $V_{\alpha\beta}= 4\epsilon_{\alpha\beta}[(\sigma_{\alpha\beta} / r)^{12}-(\sigma_{\alpha\beta} / r)^{6}]$. In the KA mixture, $\epsilon_{BB}=0.5$, $\epsilon_{AB}=1.5$, $\sigma_{BB}=0.88$ and $\sigma_{AB}=0.88$, while $\epsilon_{AA}$ and $\sigma_{AA}$ set the unit of energy and length. In the Wahnstrom mixture, $\epsilon_{AA}=\epsilon_{AB}=\epsilon_{BB}$,  $\sigma_{BB}=5/6$ and $\sigma_{AB}=11/12$, and the mass of the A type particles is twice the mass of the B type particles. 
In the mixtures, we use a version of the LJ potential that smoothly interpolates to zero between $r=2.2$ and $2.5$; in the polymer, we use a simple shifted LJ potential truncated at $r=2.25$.

Datasets for the thermal activation study are created by following established preparation protocols from the literature. For the polymer glass, we perform a slow quench from a melt temperature $T=1$ to a temperature $T=0.47$ over 100,000 LJ-times in the NPT ensemble with a pressure $p=1$. This thermodynamic path creates a supercooled fluid just above the mode coupling temperature at a reduced density of 1.04 \cite{bennemann1998molecular}. For the binary mixures, we perform a slow quench from a melt temperature $T=1$ to $T=0.44$ (KA-mixture) and $T=0.58$ (Wahnstrom-mixture) over 50,000 LJ-times in the NVT ensemble at fixed densities 1.2 (KA-mixture ) and 1.3 (Wahnstrom-mixture),  followed by two further equilibration steps of the same duration at the target temperatures in the NPT and NVT ensembles \cite{bapst2020unveiling}. This procedure ensures that the supercooled fluids are well equilibrated. Dynamics is then simulated in the microcanonical (NVE) ensemble for a time of 1,000 (polymer),  5,000 (KA model) and 1,000 (Wahnstrom model) LJ-times, which corresponds approximately to the main structural relaxation time $\tau_\alpha$ at the target temperatures for these systems \cite{bennemann1998molecular,bapst2020unveiling,royall2015strong}.

For the mechanical activation study, we start from the same melt states, but perform a very rapid quench to $T=0$ in the NVE ensemble over 500 LJ-times. We then deform the system in simple shear using the athermal quasistatic (AQS) protocol, where after each shear strain increment of $\delta \gamma = 5\times 10^{-5}$, the potential energy is minimized. Data is only collected after an initial deformation to a strain of 10\%, which erases the memory of the preparation protocol. All simulations are performed with the LAMMPS package \cite{LAMMPS}.

As a common diagnostic of local particle rearrangements, we use the well-known $D^2_{min}(i)$-measure \cite{falk1998dynamics}, which describes the nonaffine residual deformation in a spherical region around particle $i$ relative to a reference configuration. This quantity is known as an excellent indicator of irreversible plastic events, but it can also be used in the absence of global deformation. For mechanically activated events, the reference configurations are separated by $\delta \gamma$, while for thermally activated events, they are separated by $\tau_\alpha$. 

In prior studies, it has proven useful to assess structural heterogeneity from thermal trajectories in the isoconfigurational ensemble, which removes any dependence on the initial velocity distribution. Here we follow the same strategy by averaging over 30 simulations started with different velocity distributions from the same particle positions. When the squared or absolute displacement of particle $i$ is used in this average, this quantity has been termed {\em dynamic propensity} \cite{widmer2006predicting}. We find that an isoconfigurational average of $D^2_{min}(i)$ correlates to 95\% with the propensity, thus captures essentially the same information. In the athermal shear case, such an isoconfigurational averaging is not possible; instead we collect statistics over 30 disorder realizations.

\section{Results and Discussion}
\begin{figure}[t]
\begin{center}
  \includegraphics[width=6.5in]{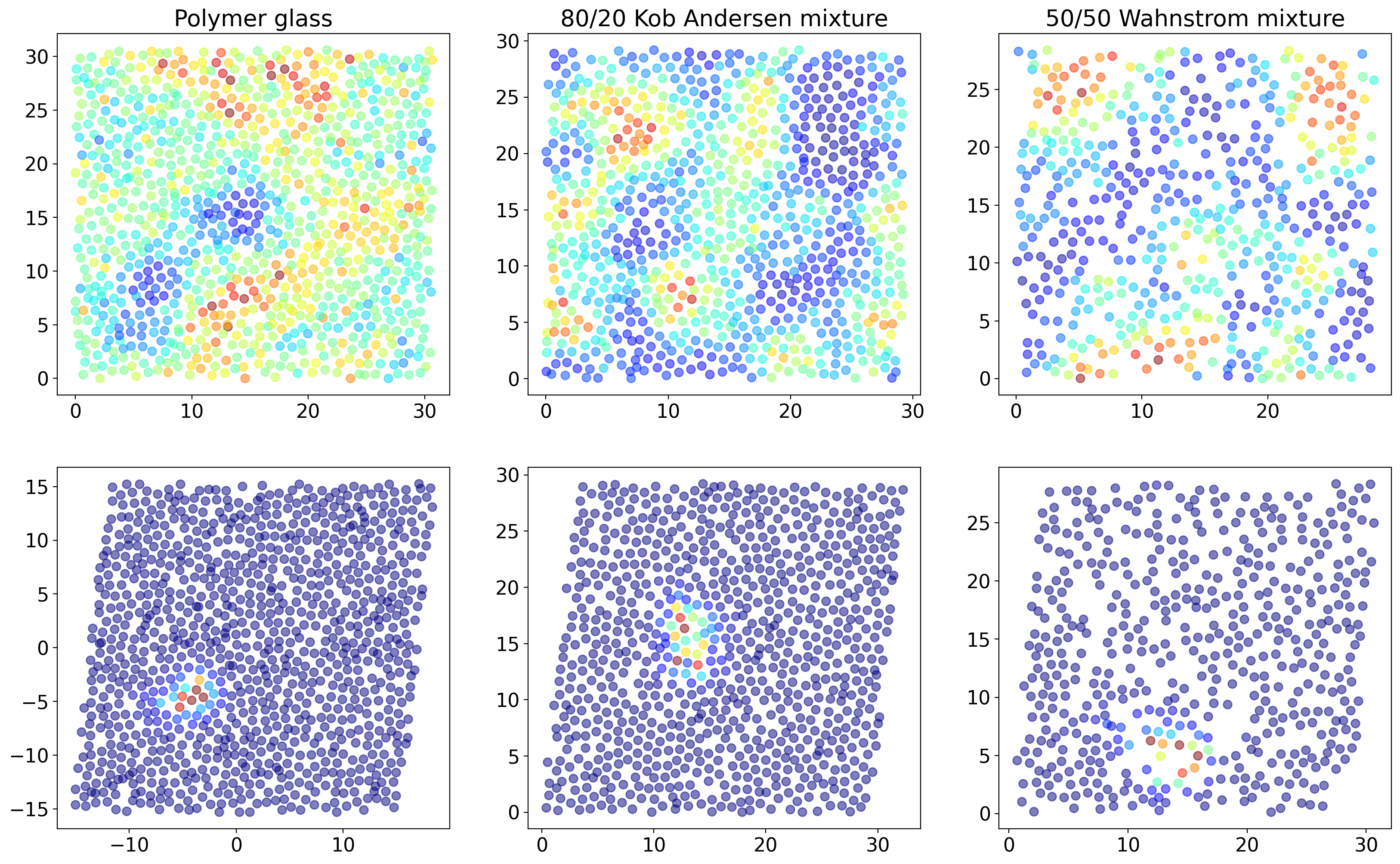}
  \caption{Example configurations of thermal excitations (top row) and a mechanically activated plastic event (bottom row) in the three model systems. Only majority (type A) particles are shown in a thin slice through the 3d simulation cell. Color indicates the magnitude of the $D^2_{min}(i)$ measure, isconfigurationally averaged for the top row (see text).}
  \label{fgr:configs}
  \end{center}
\end{figure}

We begin with a qualitative illustration of the phenomena we seek to analyze with the ACE classifier. First, we show isoconfigurationally averaged maps of $D^2_{min}(i)$ in a thin slice of one particle diameter thicknesss in the top row of Fig.~\ref{fgr:configs}. The systems are quiescent at their reference temperature, and motion occurs only through thermal activation. Only majority particles are shown, and the color reflects the amplitude of the $D^2_{min}(i)$-indicator. These maps reveal a typical picture of dynamical heterogeneity, with mobile regions (red) embedded in immobile regions (blue) when observed over a time of approximately~$\tau_\alpha$. Examples of typical single shear transformations during steady state flow of the glasses at $T=0$ are shown in the bottom row of Figure \ref{fgr:configs}. Here only a small number of particles near the core of the shear transformation are activated in an otherwise homogeneously transforming background. 

From these simulations, we construct two datasets for further analysis with an ACE classifier as follows: In the shear simulations, we define majority (type A) particles as plastically active if their $D^2_{min}$-value exceeds 0.6, and collect such events over strains of 1\% in 30 different samples. In order to achieve a balanced data set, we then add the same number of inactive particles, defined by having experienced the lowest $D^2_{min}$-values during the same strain interval. We also remove any events in which large collective avalanches have occurred, because those cannot be expected to correlate with local structure. In the thermal simulations, we simply take 5\% each of the most mobile and immobile type A particles, thus mirroring the balanced set of the mechanical simulations. ACE descriptors are then computed on the inherent structures of the initial configuration instead of the instantaneous particle positions.

Our primary objective is to characterize the structural differences in the environment of active versus passive particles with the help of the ACE classifier \eqref{eq:classifier}. To this end, we use logistic regression with Tychonov regularization to build a simple linear classifier. A linear support vector machine yields nearly identical results. We explore descriptors up to correlation order $N=3$, i.e.~up to 4-body terms. For the mono-atomic polymer, we use a maximum radial polynomial order $n_t=20$ and a total degree $d=18$ for polymers, while we set $n_t=10$ and $d=10$ for the binary mixtures. The latter means that we consider all combinations of radial order $n_t$ and $l_t$ that satisfy $\sum_{t=1}^3 n_t+w_l l_t \le d$ (here use the recommended\cite{witt2023} default of $w_l=1.5$; but other choices yield very similar results). Since we constructed balanced data sets, accuracy (i.e.~fraction of correct predictions) is a suitable performance metric for our binary classifier. Learning curves shown in the supplemental figure \ref{fgr:learning} demonstrate that all datasets are large enough to reach well-converged predictions.

\begin{figure}[t]
  \includegraphics[width=6.5in]{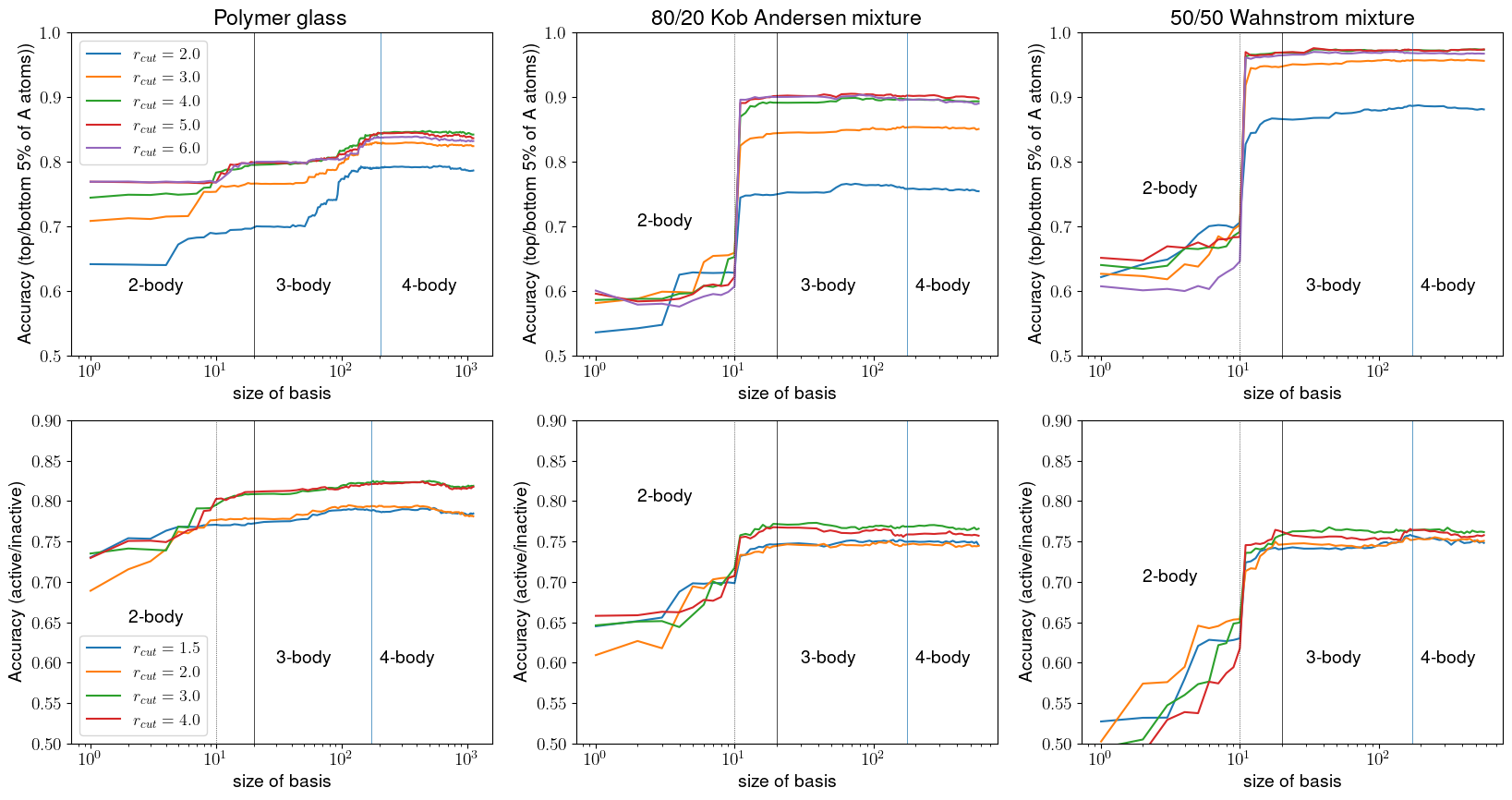}  
  \caption{Accuracy scores for binary classification using logistic regression for thermal excitations (top row) and mechanical excitations (bottom row).}
  \label{fgr:class}
\end{figure}

Figure~\ref{fgr:class} reports cross-validated accuracy scores for the thermal and mechanical activation datasets versus the number of features used by the classifier for several different values of the cutoff radius $r_{cut}$ used to compute said features. These features are sorted by body order, i.e. first only pairwise (2-body) terms are considered, then 3-body terms are added, and finally 4-body terms are included as well. Vertical lines indicate when a higher body order becomes part of the basis set. For the binary mixtures, the vertical dashed line demarcates the border between radial features linking atoms of the same type (A to A) and opposite type (A to B).

Analyzing first the thermal activation mechanism in the top row, we find a strong dependence of the accuracy on $r_{cut}$, with values saturating at $r_{cut}=4$ for all models. This large value suggests that structure over multiple neighbor shells is important for determining the local mobility. In general, the accuracy increases with increasing basis size, but different behavior is found in the polymer vs. the binary mixtures. In the polymer, using the first feature alone allows for classification with 75\% accuracy. This value  reaches 80\% once 20 radial descriptors are taken into account, and further increases to 85\% with the help of 3-body terms. In the binary mixtures by contrast, only modest accuracy of 60-65\% is achieved with the first 10 radial features, which describe pairwise interactions between the central particle and neighboring particles of the same type, i.e. both particles are of type A. However, the accuracy dramatically increases to 90\% (KA-mixture) and 97\% (Wahnstrom-mixture) as soon as some information about correlations between the central atom and B type neighbors are included in the classifier. The thermal mobility is therefore strongly affected by the configuration of the smaller B particles in the local environment. Unlike the polymer system, including 3- and 4-body terms does not further increase the accuracy substantially; there is only a very modest rise once 3-body terms involving interactions with B type particles start to be included.  

Yet another scenario unfolds when shear transformations activate the particles, as can be appreciated in the bottom row of Fig.~\ref{fgr:class}. Our first observation is that the accuracy varies much less with $r_{cut}$ and saturates at a lower value of $r_{cut}=3.0$. This means that mechanical activation is mostly sensitive to short range structure in the first two neighbor shells. In the polymer, the accuracy rises from 73\% with just one feature to 83\% when all 3-body terms are included. In the KA-mixture, we see that an accuracy of 70\% is reached after 10 basis functions have been included in the model, which suggests that now it is the 2-body correlations with the majority A particles that are most important. Adding in correlations with B particles then achieves maximum accuracy, and, as in the thermal case, higher order body order correlations do not appear to play any significant role. Interestingly, the Wahnstrom mixture behaves differently, and all radial basis functions are needed to reach optimal accuracy. As in the thermal case, higher order body correlations do not improve the model. The overall lower accuracy scores for the classification of mechanically activated particles are primarily due to the lack of the isoconfigurational ensemble, combined with the more collective nature of the shear transformations. Supplemental figure ~\ref{fgr:regression} shows results from linear regression on the same dataset using the Pearson correlation coefficient as an alternative metric, which confirms that the trends observed in figure \ref{fgr:class} are not specific to our choice of treating the data as a classification problem.
 
\begin{figure}[t]
  \includegraphics[width=6cm]{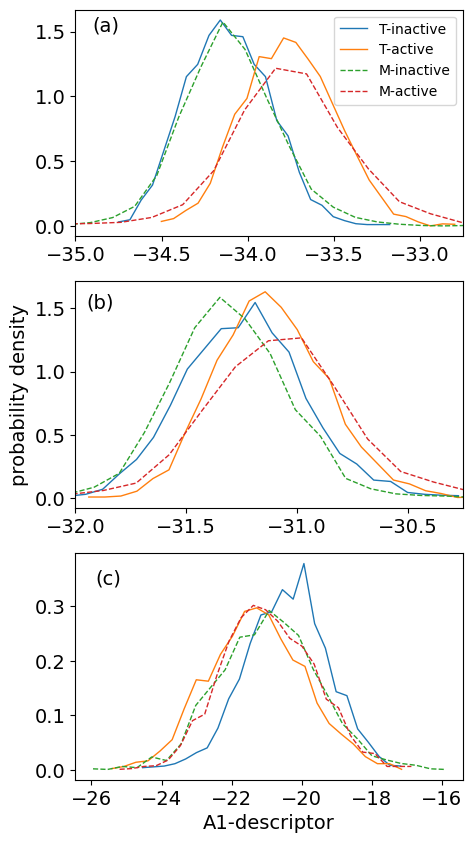}  
  \caption{Distributions of first radial descriptor for (a) the polymer system, (b) the KA-mixture, and (c) the Wahnstrom-mixture. All descriptors were computed with $r_c=4$. Solid lines refer to data from thermal (T) excitation, while dashed lines show distributions from mechanical (M) excitation.}
   \label{fgr:dists}
\end{figure}

The results of figure \ref{fgr:class} fall in line with many previous studies that demonstrate strong links between local structure, relaxation and plastic events in glasses. The correlation varies quite strongly between models \cite{paret2020assessing,jung2023roadmap}, and different aspects of the structure are important for different relaxation mechanisms. Further insight into this aspect can be gained from a detailed inspection of the distribution of the ACE descriptors. In figure \ref{fgr:dists}, we examine the distributions of the first $(n=1)$ radial descriptor separated into active and inactive particles for both thermal and mechanical activation. For both excitation mechanisms in the polymer, the two distributions are well separated, which explains why just using this one radial feature already allows for classification with 74\% accuracy. A similar behavior can be seen in the KA-model. We also see that the magnitude of the mean of the descriptor value is larger for the inactive particles, consistent with the notion that their local environment is denser. 

For thermal excitation in the Wahnstrom model, the descriptor also discriminates between active and inactive particles, but now the active particles have a larger magnitude of the mean value due to their larger mass. For mechanical excitation the distributions overlap, which explains the low scores of the classifier for mechanical activation when only this feature is used. 

\begin{figure}[t]
  \includegraphics[width=6.5in]{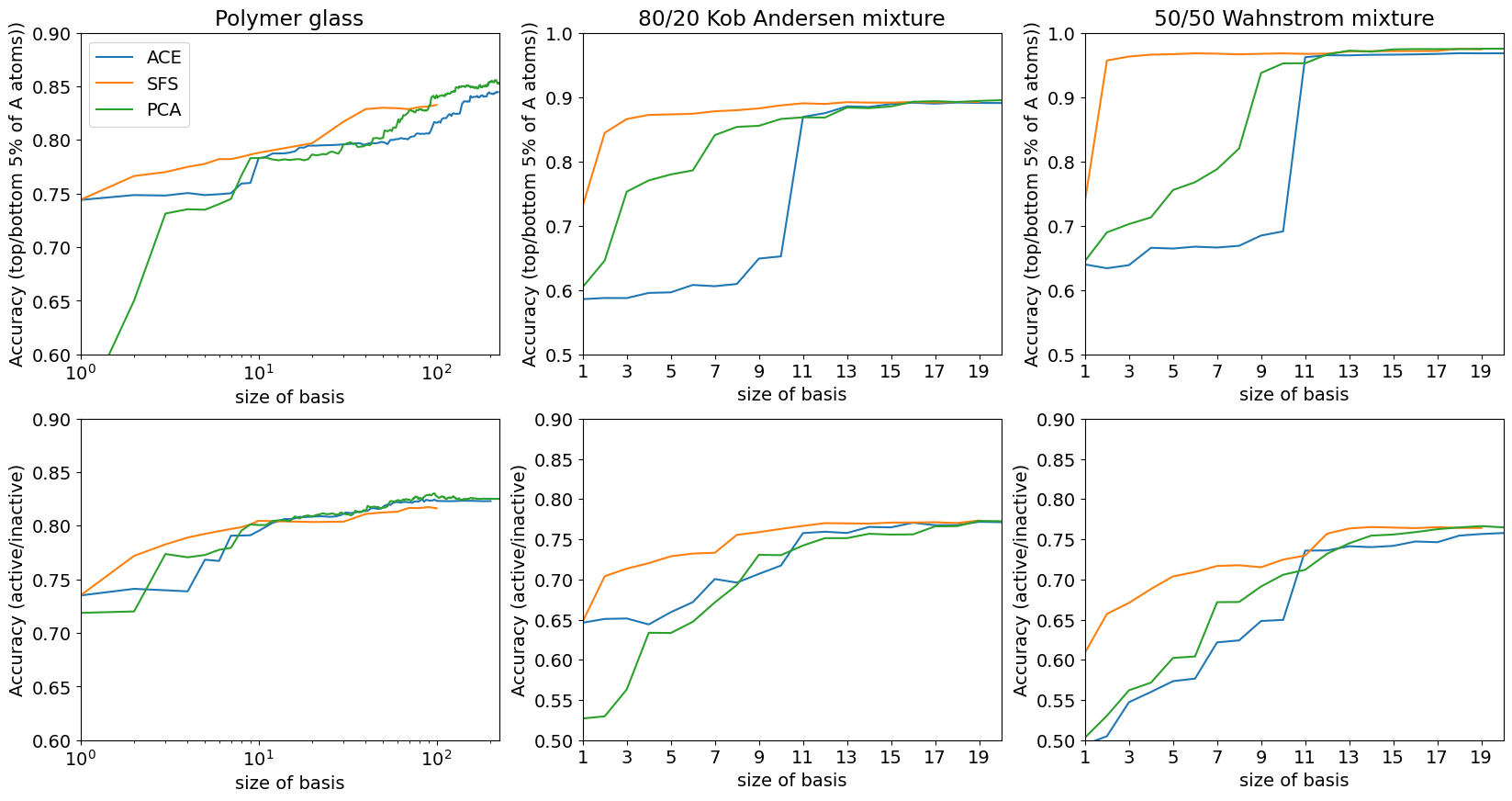}  
  \caption{Feature selection for thermal excitation (top row) and  mechanical excitation (bottom row) with sequential feature selection (SFS) and principal component analysis (PCA). Only 2- and 3-body terms are considered.}
  \label{fgr:featsel}
\end{figure}

Figure \ref{fgr:dists} shows that for building optimal classifiers, different structural feature sets should be used for different target properties. The structural features that are most important for predicting the target quantity can be obtained with sequential feature selection. In this technique, a feature subset is built by sequentially adding at each step the best feature that most improves the (cross-validated) score. This construction results in an optimal basis set of a given size. The top row of Figure \ref{fgr:featsel} compares the accuracy score of such an optimally selected basis to a basis in standard ACE order (as was used in Fig.~\ref{fgr:class}). In the polymer case, convergence can be accelerated but the number of features required to reach maximum accuracy cannot be reduced. Both for thermal and mechanical activation, the first radial term in the ACE expansion $(n=1)$ is also the most important feature selected  (note that the first 3-body term $(n_1=n_2=1, l_1=l_2=0)$ generates an almost identical score). A different and more dramatic result is found in the thermal excitation of binary mixtures, where the accuracy score converges very rapidly after just a few features have been included. In both mixtures, the first two features selected are the first order $(n=1)$ two-body radial terms B1 and A1 between B and A particles, resp. For mechanical excitation (bottom row of Figure \ref{fgr:featsel}), the prediction converges more slowly, and about 10 basis functions are needed for optimal performance. In this case, the first two basis functions selected are A1 and B1 for the KA mixture and A8 and A7 for the Wahnstrom mixture.

Another popular method for dimensionality reduction of a data set is principal component analysis (PCA), which performs a linear transformation of the data so that the first principal components maximize the explained variance. However, Figure \ref{fgr:featsel} shows that classifiers using PCA-ordered components instead of bare basis functions do not necessarily converge faster than those using the standard order of basis functions. Only the polymer case shows improvement.

\section{Conclusions}
We have analyzed the correlation between purely geometric aspects of local structure and local mechanical and thermal excitations in several model glass formers as a linear classification problem, using features generated with the atomic cluster expansion (ACE). Our results afford the following conclusions: 
(i) Expanding on previous reports \cite{paret2020assessing,jung2023predicting}, the degree of correlation between structure and dynamics can vary substantially between glass forming models. 
(ii) In binary mixtures, only pairwise (2-body) correlations are relevant for correlating local structure and dynamics, while higher order multibody terms do not appear to be important. This result appears in qualitative agreement with the work of Boattini et al.~\cite{boattini2021}, who also found most of the correlations coming already from 2-body terms in their set of descriptors. It is also consistent with the sucesss of the GlassMLP model proposed by Jung et al. \cite{jung2023predicting,jung2024dynamic}, which uses a 2-body correlation (coarse-grained local density with an exponential weighting function) as one of the features.  
Both KA and Wahnstrom mixtures exhibit well-known locally favored high-symmetry structures (bicapped square antiprism or icosahedron) \cite{royall2015strong}. The local environment in the polymer model lacks these features, requiring 3-body terms for maximum performance.  
(iii) While thermal excitations are sensitive to structural features up to four particle diameters, most of the signal for mechanical excitations is already obtained at a shorter range of three particle diameters. This suggests that it is primarily the structure in the first two nearest neighbor shells that control local mechanical instability. The ability of particles to leave local cages via thermal activation, however, is influenced by medium-range packing order in subsequent shells as well. (iv) In binary mixtures, thermal excitation of the majority type A particles is much more strongly influenced by the arrangement of smaller B type particles than the triggering of shear transformations. The extremely fast saturation of the accuracy with increasing basis size in the binary mixtures suggests that these are actually not the best choice for testing ML based structural classifiers, while the polymer glass provides a more challenging test case. 

Although the linear classifiers we constructed separate particles that are likely to rearrange from those that are not, they do this for different underlying structural reasons.  Remarkably, the radial basis of ACE allows for the construction of minimal classifiers that yield acceptable predictions already with very few structural features. Recalling that the radial basis can be interpreted as coarse-grained local densities, this highlights their role as most important structural order parameter. We suggest that the systematic and robust set of structural features generated by ACE are ideally suited for future exploration of more complex glass formers such as network glasses, multicomponent metallic glasses, or polydisperse colloidal mixtures. Given the documented ability of deep learning to improve predictions from linear models, a promising direction appears to be the application of MACE, the equivariant message-passing extension of the ACE formalism, to the prediction of local glassy dynamics \cite{batatia2022mace}.

\begin{acknowledgement}
We thank Gerhard Jung for insightful discussions and a critical reading of the manuscript. JR would like to thank Professor Mark Ediger for many intensive and productive scientific discussions over the past 20 years, and in particular his interest in studies of the mechanical behavior of polymer glasses with molecular simulations.

\end{acknowledgement}


\newpage
\begin{suppinfo}

\begin{figure}[h]
  \includegraphics[width=7cm]{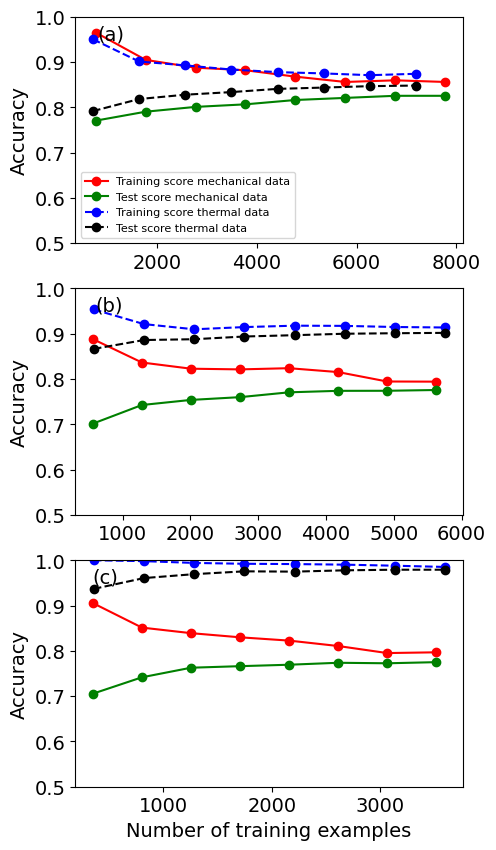}  
  \caption{Learning curves for (a) the polymer system, (b) the KA-mixture, and (c) the Wahnstrom-mixture.}
  \label{fgr:learning}
\end{figure}

\begin{figure}[h]
  \includegraphics[width=6.5in]{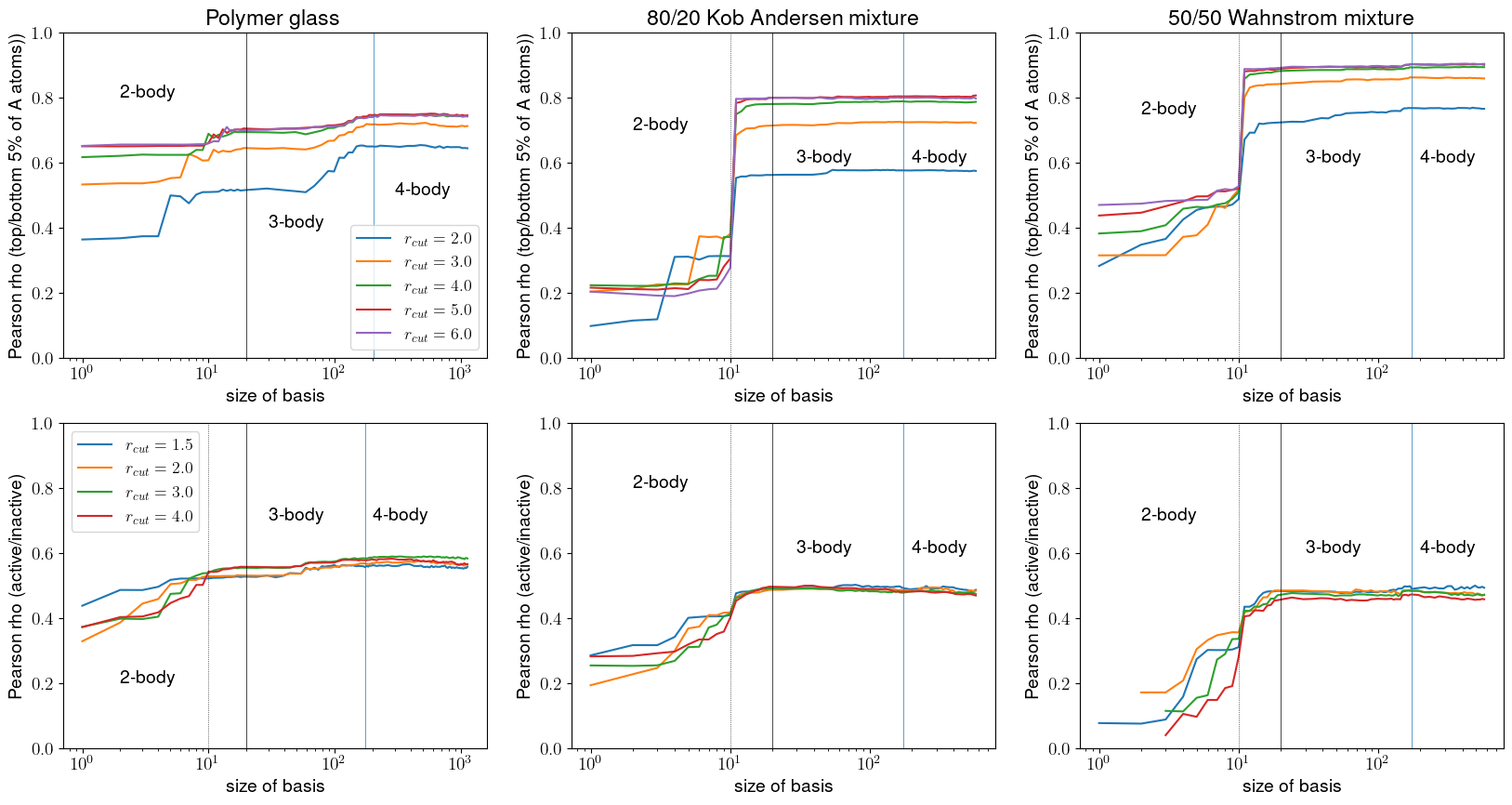}  
  \caption{Pearson correlation coefficient obtaind from Bayesian ARD (Atomatic Relevance Detection) regression for thermal excitations (top row) and mechanical excitations (bottom row). Learning curves for (a) the polymer system, (b) the KA-mixture, and (c) the Wahnstrom-mixture. The dataset is the same as the one studied in Fig.~\ref{fgr:class}, with the only difference being that the target variable $D^2_{min}(i)$ is not binarized.}
  \label{fgr:regression}
\end{figure}

\end{suppinfo}

\newpage
\bibliography{achemso-demo}

\providecommand{\latin}[1]{#1}
\makeatletter
\providecommand{\doi}
  {\begingroup\let\do\@makeother\dospecials
  \catcode`\{=1 \catcode`\}=2 \doi@aux}
\providecommand{\doi@aux}[1]{\endgroup\texttt{#1}}
\makeatother
\providecommand*\mcitethebibliography{\thebibliography}
\csname @ifundefined\endcsname{endmcitethebibliography}  {\let\endmcitethebibliography\endthebibliography}{}
\begin{mcitethebibliography}{36}
\providecommand*\natexlab[1]{#1}
\providecommand*\mciteSetBstSublistMode[1]{}
\providecommand*\mciteSetBstMaxWidthForm[2]{}
\providecommand*\mciteBstWouldAddEndPuncttrue
  {\def\EndOfBibitem{\unskip.}}
\providecommand*\mciteBstWouldAddEndPunctfalse
  {\let\EndOfBibitem\relax}
\providecommand*\mciteSetBstMidEndSepPunct[3]{}
\providecommand*\mciteSetBstSublistLabelBeginEnd[3]{}
\providecommand*\EndOfBibitem{}
\mciteSetBstSublistMode{f}
\mciteSetBstMaxWidthForm{subitem}{(\alph{mcitesubitemcount})}
\mciteSetBstSublistLabelBeginEnd
  {\mcitemaxwidthsubitemform\space}
  {\relax}
  {\relax}

\bibitem[Ediger(2000)]{ediger2000spatially}
Ediger,~M.~D. Spatially heterogeneous dynamics in supercooled liquids. \emph{Annual review of physical chemistry} \textbf{2000}, \emph{51}, 99--128\relax
\mciteBstWouldAddEndPuncttrue
\mciteSetBstMidEndSepPunct{\mcitedefaultmidpunct}
{\mcitedefaultendpunct}{\mcitedefaultseppunct}\relax
\EndOfBibitem
\bibitem[Berthier \latin{et~al.}(2011)Berthier, Biroli, Bouchaud, Cipelletti, and van Saarloos]{berthier2011dynamical}
Berthier,~L.; Biroli,~G.; Bouchaud,~J.-P.; Cipelletti,~L.; van Saarloos,~W. \emph{Dynamical heterogeneities in glasses, colloids, and granular media}; OUP Oxford, 2011; Vol. 150\relax
\mciteBstWouldAddEndPuncttrue
\mciteSetBstMidEndSepPunct{\mcitedefaultmidpunct}
{\mcitedefaultendpunct}{\mcitedefaultseppunct}\relax
\EndOfBibitem
\bibitem[Argon(1979)]{argon1979plastic}
Argon,~A. Plastic deformation in metallic glasses. \emph{Acta metallurgica} \textbf{1979}, \emph{27}, 47--58\relax
\mciteBstWouldAddEndPuncttrue
\mciteSetBstMidEndSepPunct{\mcitedefaultmidpunct}
{\mcitedefaultendpunct}{\mcitedefaultseppunct}\relax
\EndOfBibitem
\bibitem[Richard \latin{et~al.}(2020)Richard, Ozawa, Patinet, Stanifer, Shang, Ridout, Xu, Zhang, Morse, Barrat, \latin{et~al.} others]{richard2020predicting}
Richard,~D.; Ozawa,~M.; Patinet,~S.; Stanifer,~E.; Shang,~B.; Ridout,~S.; Xu,~B.; Zhang,~G.; Morse,~P.; Barrat,~J.-L.; others Predicting plasticity in disordered solids from structural indicators. \emph{Physical Review Materials} \textbf{2020}, \emph{4}, 113609\relax
\mciteBstWouldAddEndPuncttrue
\mciteSetBstMidEndSepPunct{\mcitedefaultmidpunct}
{\mcitedefaultendpunct}{\mcitedefaultseppunct}\relax
\EndOfBibitem
\bibitem[Behler and Parrinello(2007)Behler, and Parrinello]{behler2007generalized}
Behler,~J.; Parrinello,~M. Generalized neural-network representation of high-dimensional potential-energy surfaces. \emph{Physical review letters} \textbf{2007}, \emph{98}, 146401\relax
\mciteBstWouldAddEndPuncttrue
\mciteSetBstMidEndSepPunct{\mcitedefaultmidpunct}
{\mcitedefaultendpunct}{\mcitedefaultseppunct}\relax
\EndOfBibitem
\bibitem[Bart\'ok \latin{et~al.}(2013)Bart\'ok, Kondor, and Cs\'anyi]{bartok2013}
Bart\'ok,~A.~P.; Kondor,~R.; Cs\'anyi,~G. On representing chemical environments. \emph{Phys. Rev. B} \textbf{2013}, \emph{87}, 184115\relax
\mciteBstWouldAddEndPuncttrue
\mciteSetBstMidEndSepPunct{\mcitedefaultmidpunct}
{\mcitedefaultendpunct}{\mcitedefaultseppunct}\relax
\EndOfBibitem
\bibitem[Jung \latin{et~al.}(2023)Jung, Alkemade, Bapst, Coslovich, Filion, Landes, Liu, Pezzicoli, Shiba, Volpe, \latin{et~al.} others]{jung2023roadmap}
Jung,~G.; Alkemade,~R.~M.; Bapst,~V.; Coslovich,~D.; Filion,~L.; Landes,~F.~P.; Liu,~A.; Pezzicoli,~F.~S.; Shiba,~H.; Volpe,~G.; others Roadmap on machine learning glassy liquids. \emph{arXiv preprint arXiv:2311.14752} \textbf{2023}, \relax
\mciteBstWouldAddEndPunctfalse
\mciteSetBstMidEndSepPunct{\mcitedefaultmidpunct}
{}{\mcitedefaultseppunct}\relax
\EndOfBibitem
\bibitem[Drautz(2019)]{drautz2019}
Drautz,~R. Atomic cluster expansion for accurate and transferable interatomic potentials. \emph{Phys. Rev. B} \textbf{2019}, \emph{99}, 014104\relax
\mciteBstWouldAddEndPuncttrue
\mciteSetBstMidEndSepPunct{\mcitedefaultmidpunct}
{\mcitedefaultendpunct}{\mcitedefaultseppunct}\relax
\EndOfBibitem
\bibitem[Dusson \latin{et~al.}(2022)Dusson, Bachmayr, Csányi, Drautz, Etter, {van der Oord}, and Ortner]{dusson2022}
Dusson,~G.; Bachmayr,~M.; Csányi,~G.; Drautz,~R.; Etter,~S.; {van der Oord},~C.; Ortner,~C. Atomic cluster expansion: Completeness, efficiency and stability. \emph{Journal of Computational Physics} \textbf{2022}, \emph{454}, 110946\relax
\mciteBstWouldAddEndPuncttrue
\mciteSetBstMidEndSepPunct{\mcitedefaultmidpunct}
{\mcitedefaultendpunct}{\mcitedefaultseppunct}\relax
\EndOfBibitem
\bibitem[Cubuk \latin{et~al.}(2015)Cubuk, Schoenholz, Rieser, Malone, Rottler, Durian, Kaxiras, and Liu]{cubuk2015}
Cubuk,~E.~D.; Schoenholz,~S.~S.; Rieser,~J.~M.; Malone,~B.~D.; Rottler,~J.; Durian,~D.~J.; Kaxiras,~E.; Liu,~A.~J. Identifying Structural Flow Defects in Disordered Solids Using Machine-Learning Methods. \emph{Phys. Rev. Lett.} \textbf{2015}, \emph{114}, 108001\relax
\mciteBstWouldAddEndPuncttrue
\mciteSetBstMidEndSepPunct{\mcitedefaultmidpunct}
{\mcitedefaultendpunct}{\mcitedefaultseppunct}\relax
\EndOfBibitem
\bibitem[Schoenholz \latin{et~al.}(2016)Schoenholz, Cubuk, Sussman, Kaxiras, and Liu]{schoenholz2016structural}
Schoenholz,~S.~S.; Cubuk,~E.~D.; Sussman,~D.~M.; Kaxiras,~E.; Liu,~A.~J. A structural approach to relaxation in glassy liquids. \emph{Nature Physics} \textbf{2016}, \emph{12}, 469--471\relax
\mciteBstWouldAddEndPuncttrue
\mciteSetBstMidEndSepPunct{\mcitedefaultmidpunct}
{\mcitedefaultendpunct}{\mcitedefaultseppunct}\relax
\EndOfBibitem
\bibitem[Wu \latin{et~al.}(2023)Wu, Xu, Zhang, and Guan]{wu2023machine}
Wu,~Y.; Xu,~B.; Zhang,~X.; Guan,~P. Machine-learning inspired density-fluctuation model of local structural instability in metallic glasses. \emph{Acta Materialia} \textbf{2023}, \emph{247}, 118741\relax
\mciteBstWouldAddEndPuncttrue
\mciteSetBstMidEndSepPunct{\mcitedefaultmidpunct}
{\mcitedefaultendpunct}{\mcitedefaultseppunct}\relax
\EndOfBibitem
\bibitem[Boattini \latin{et~al.}(2021)Boattini, Smallenburg, and Filion]{boattini2021}
Boattini,~E.; Smallenburg,~F.; Filion,~L. Averaging Local Structure to Predict the Dynamic Propensity in Supercooled Liquids. \emph{Phys. Rev. Lett.} \textbf{2021}, \emph{127}, 088007\relax
\mciteBstWouldAddEndPuncttrue
\mciteSetBstMidEndSepPunct{\mcitedefaultmidpunct}
{\mcitedefaultendpunct}{\mcitedefaultseppunct}\relax
\EndOfBibitem
\bibitem[Alkemade \latin{et~al.}(2023)Alkemade, Smallenburg, and Filion]{alkemade2023improving}
Alkemade,~R.~M.; Smallenburg,~F.; Filion,~L. Improving the prediction of glassy dynamics by pinpointing the local cage. \emph{The Journal of Chemical Physics} \textbf{2023}, \emph{158}\relax
\mciteBstWouldAddEndPuncttrue
\mciteSetBstMidEndSepPunct{\mcitedefaultmidpunct}
{\mcitedefaultendpunct}{\mcitedefaultseppunct}\relax
\EndOfBibitem
\bibitem[Wang \latin{et~al.}(2020)Wang, Ding, Zhang, Podryabinkin, Shapeev, and Ma]{wang2020predicting}
Wang,~Q.; Ding,~J.; Zhang,~L.; Podryabinkin,~E.; Shapeev,~A.; Ma,~E. Predicting the propensity for thermally activated $\beta$ events in metallic glasses via interpretable machine learning. \emph{npj Computational Materials} \textbf{2020}, \emph{6}, 194\relax
\mciteBstWouldAddEndPuncttrue
\mciteSetBstMidEndSepPunct{\mcitedefaultmidpunct}
{\mcitedefaultendpunct}{\mcitedefaultseppunct}\relax
\EndOfBibitem
\bibitem[Coslovich \latin{et~al.}(2022)Coslovich, Jack, and Paret]{coslovich2022}
Coslovich,~D.; Jack,~R.~L.; Paret,~J. {Dimensionality reduction of local structure in glassy binary mixtures}. \emph{The Journal of Chemical Physics} \textbf{2022}, \emph{157}, 204503\relax
\mciteBstWouldAddEndPuncttrue
\mciteSetBstMidEndSepPunct{\mcitedefaultmidpunct}
{\mcitedefaultendpunct}{\mcitedefaultseppunct}\relax
\EndOfBibitem
\bibitem[Bapst \latin{et~al.}(2020)Bapst, Keck, Grabska-Barwi{\'n}ska, Donner, Cubuk, Schoenholz, Obika, Nelson, Back, Hassabis, \latin{et~al.} others]{bapst2020unveiling}
Bapst,~V.; Keck,~T.; Grabska-Barwi{\'n}ska,~A.; Donner,~C.; Cubuk,~E.~D.; Schoenholz,~S.~S.; Obika,~A.; Nelson,~A.~W.; Back,~T.; Hassabis,~D.; others Unveiling the predictive power of static structure in glassy systems. \emph{Nature physics} \textbf{2020}, \emph{16}, 448--454\relax
\mciteBstWouldAddEndPuncttrue
\mciteSetBstMidEndSepPunct{\mcitedefaultmidpunct}
{\mcitedefaultendpunct}{\mcitedefaultseppunct}\relax
\EndOfBibitem
\bibitem[Fan and Ma(2021)Fan, and Ma]{fan2021predicting}
Fan,~Z.; Ma,~E. Predicting orientation-dependent plastic susceptibility from static structure in amorphous solids via deep learning. \emph{Nature communications} \textbf{2021}, \emph{12}, 1506\relax
\mciteBstWouldAddEndPuncttrue
\mciteSetBstMidEndSepPunct{\mcitedefaultmidpunct}
{\mcitedefaultendpunct}{\mcitedefaultseppunct}\relax
\EndOfBibitem
\bibitem[Shiba \latin{et~al.}(2023)Shiba, Hanai, Suzumura, and Shimokawabe]{shiba2023}
Shiba,~H.; Hanai,~M.; Suzumura,~T.; Shimokawabe,~T. {BOTAN: BOnd TArgeting Network for prediction of slow glassy dynamics by machine learning relative motion}. \emph{The Journal of Chemical Physics} \textbf{2023}, \emph{158}, 084503\relax
\mciteBstWouldAddEndPuncttrue
\mciteSetBstMidEndSepPunct{\mcitedefaultmidpunct}
{\mcitedefaultendpunct}{\mcitedefaultseppunct}\relax
\EndOfBibitem
\bibitem[Jung \latin{et~al.}(2023)Jung, Biroli, and Berthier]{jung2023predicting}
Jung,~G.; Biroli,~G.; Berthier,~L. Predicting dynamic heterogeneity in glass-forming liquids by physics-inspired machine learning. \emph{Physical Review Letters} \textbf{2023}, \emph{130}, 238202\relax
\mciteBstWouldAddEndPuncttrue
\mciteSetBstMidEndSepPunct{\mcitedefaultmidpunct}
{\mcitedefaultendpunct}{\mcitedefaultseppunct}\relax
\EndOfBibitem
\bibitem[Oyama \latin{et~al.}(2023)Oyama, Koyama, and Kawasaki]{oyama2023}
Oyama,~N.; Koyama,~S.; Kawasaki,~T. What do deep neural networks find in disordered structures of glasses? \emph{Frontiers in Physics} \textbf{2023}, \emph{10}\relax
\mciteBstWouldAddEndPuncttrue
\mciteSetBstMidEndSepPunct{\mcitedefaultmidpunct}
{\mcitedefaultendpunct}{\mcitedefaultseppunct}\relax
\EndOfBibitem
\bibitem[Pezzicoli \latin{et~al.}(2024)Pezzicoli, Charpiat, and Landes]{pezzicoli2024}
Pezzicoli,~F.~S.; Charpiat,~G.; Landes,~F.~P. {Rotation-equivariant graph neural networks for learning glassy liquids representations}. \emph{SciPost Phys.} \textbf{2024}, \emph{16}, 136\relax
\mciteBstWouldAddEndPuncttrue
\mciteSetBstMidEndSepPunct{\mcitedefaultmidpunct}
{\mcitedefaultendpunct}{\mcitedefaultseppunct}\relax
\EndOfBibitem
\bibitem[Jung \latin{et~al.}(2024)Jung, Biroli, and Berthier]{jung2024dynamic}
Jung,~G.; Biroli,~G.; Berthier,~L. Dynamic heterogeneity at the experimental glass transition predicted by transferable machine learning. \emph{Phys. Rev. B} \textbf{2024}, \emph{109}, 064205\relax
\mciteBstWouldAddEndPuncttrue
\mciteSetBstMidEndSepPunct{\mcitedefaultmidpunct}
{\mcitedefaultendpunct}{\mcitedefaultseppunct}\relax
\EndOfBibitem
\bibitem[Witt \latin{et~al.}(2023)Witt, van~der Oord, Gelžinytė, Järvinen, Ross, Darby, Ho, Baldwin, Sachs, Kermode, Bernstein, Csányi, and Ortner]{witt2023}
Witt,~W.~C.; van~der Oord,~C.; Gelžinytė,~E.; Järvinen,~T.; Ross,~A.; Darby,~J.~P.; Ho,~C.~H.; Baldwin,~W.~J.; Sachs,~M.; Kermode,~J.; Bernstein,~N.; Csányi,~G.; Ortner,~C. {ACEpotentials.jl: A Julia implementation of the atomic cluster expansion}. \emph{The Journal of Chemical Physics} \textbf{2023}, \emph{159}, 164101\relax
\mciteBstWouldAddEndPuncttrue
\mciteSetBstMidEndSepPunct{\mcitedefaultmidpunct}
{\mcitedefaultendpunct}{\mcitedefaultseppunct}\relax
\EndOfBibitem
\bibitem[Ho \latin{et~al.}(2024)Ho, Gutleb, and Ortner]{pureace2024}
Ho,~C.~H.; Gutleb,~T.~S.; Ortner,~C. Atomic Cluster Expansion without Self-Interaction. \emph{J. Comp. Phys.} \textbf{2024}, \emph{515}, 113271\relax
\mciteBstWouldAddEndPuncttrue
\mciteSetBstMidEndSepPunct{\mcitedefaultmidpunct}
{\mcitedefaultendpunct}{\mcitedefaultseppunct}\relax
\EndOfBibitem
\bibitem[Kremer and Grest(1990)Kremer, and Grest]{kremer1990dynamics}
Kremer,~K.; Grest,~G.~S. Dynamics of entangled linear polymer melts: A molecular-dynamics simulation. \emph{The Journal of Chemical Physics} \textbf{1990}, \emph{92}, 5057--5086\relax
\mciteBstWouldAddEndPuncttrue
\mciteSetBstMidEndSepPunct{\mcitedefaultmidpunct}
{\mcitedefaultendpunct}{\mcitedefaultseppunct}\relax
\EndOfBibitem
\bibitem[Kob and Andersen(1995)Kob, and Andersen]{kob1995testing}
Kob,~W.; Andersen,~H.~C. Testing mode-coupling theory for a supercooled binary Lennard-Jones mixture I: The van Hove correlation function. \emph{Physical Review E} \textbf{1995}, \emph{51}, 4626\relax
\mciteBstWouldAddEndPuncttrue
\mciteSetBstMidEndSepPunct{\mcitedefaultmidpunct}
{\mcitedefaultendpunct}{\mcitedefaultseppunct}\relax
\EndOfBibitem
\bibitem[Wahnstr{\"o}m(1991)]{wahnstrom1991molecular}
Wahnstr{\"o}m,~G. Molecular-dynamics study of a supercooled two-component Lennard-Jones system. \emph{Physical Review A} \textbf{1991}, \emph{44}, 3752\relax
\mciteBstWouldAddEndPuncttrue
\mciteSetBstMidEndSepPunct{\mcitedefaultmidpunct}
{\mcitedefaultendpunct}{\mcitedefaultseppunct}\relax
\EndOfBibitem
\bibitem[Bennemann \latin{et~al.}(1998)Bennemann, Paul, Binder, and D{\"u}nweg]{bennemann1998molecular}
Bennemann,~C.; Paul,~W.; Binder,~K.; D{\"u}nweg,~B. Molecular-dynamics simulations of the thermal glass transition in polymer melts: $\alpha$-relaxation behavior. \emph{Physical Review E} \textbf{1998}, \emph{57}, 843\relax
\mciteBstWouldAddEndPuncttrue
\mciteSetBstMidEndSepPunct{\mcitedefaultmidpunct}
{\mcitedefaultendpunct}{\mcitedefaultseppunct}\relax
\EndOfBibitem
\bibitem[Royall \latin{et~al.}(2015)Royall, Malins, Dunleavy, and Pinney]{royall2015strong}
Royall,~C.~P.; Malins,~A.; Dunleavy,~A.~J.; Pinney,~R. Strong geometric frustration in model glassformers. \emph{Journal of Non-Crystalline Solids} \textbf{2015}, \emph{407}, 34--43\relax
\mciteBstWouldAddEndPuncttrue
\mciteSetBstMidEndSepPunct{\mcitedefaultmidpunct}
{\mcitedefaultendpunct}{\mcitedefaultseppunct}\relax
\EndOfBibitem
\bibitem[Thompson \latin{et~al.}(2022)Thompson, Aktulga, Berger, Bolintineanu, Brown, Crozier, in~'t Veld, Kohlmeyer, Moore, Nguyen, Shan, Stevens, Tranchida, Trott, and Plimpton]{LAMMPS}
Thompson,~A.~P.; Aktulga,~H.~M.; Berger,~R.; Bolintineanu,~D.~S.; Brown,~W.~M.; Crozier,~P.~S.; in~'t Veld,~P.~J.; Kohlmeyer,~A.; Moore,~S.~G.; Nguyen,~T.~D.; Shan,~R.; Stevens,~M.~J.; Tranchida,~J.; Trott,~C.; Plimpton,~S.~J. {LAMMPS} - a flexible simulation tool for particle-based materials modeling at the atomic, meso, and continuum scales. \emph{Comp. Phys. Comm.} \textbf{2022}, \emph{271}, 108171\relax
\mciteBstWouldAddEndPuncttrue
\mciteSetBstMidEndSepPunct{\mcitedefaultmidpunct}
{\mcitedefaultendpunct}{\mcitedefaultseppunct}\relax
\EndOfBibitem
\bibitem[Falk and Langer(1998)Falk, and Langer]{falk1998dynamics}
Falk,~M.~L.; Langer,~J.~S. Dynamics of viscoplastic deformation in amorphous solids. \emph{Physical Review E} \textbf{1998}, \emph{57}, 7192\relax
\mciteBstWouldAddEndPuncttrue
\mciteSetBstMidEndSepPunct{\mcitedefaultmidpunct}
{\mcitedefaultendpunct}{\mcitedefaultseppunct}\relax
\EndOfBibitem
\bibitem[Widmer-Cooper and Harrowell(2006)Widmer-Cooper, and Harrowell]{widmer2006predicting}
Widmer-Cooper,~A.; Harrowell,~P. Predicting the Long-Time Dynamic Heterogeneity in a Supercooled Liquid on the Basis of Short-Time Heterogeneities. \emph{Physical review letters} \textbf{2006}, \emph{96}, 185701\relax
\mciteBstWouldAddEndPuncttrue
\mciteSetBstMidEndSepPunct{\mcitedefaultmidpunct}
{\mcitedefaultendpunct}{\mcitedefaultseppunct}\relax
\EndOfBibitem
\bibitem[Paret \latin{et~al.}(2020)Paret, Jack, and Coslovich]{paret2020assessing}
Paret,~J.; Jack,~R.~L.; Coslovich,~D. Assessing the structural heterogeneity of supercooled liquids through community inference. \emph{The Journal of chemical physics} \textbf{2020}, \emph{152}\relax
\mciteBstWouldAddEndPuncttrue
\mciteSetBstMidEndSepPunct{\mcitedefaultmidpunct}
{\mcitedefaultendpunct}{\mcitedefaultseppunct}\relax
\EndOfBibitem
\bibitem[Batatia \latin{et~al.}(2022)Batatia, Kovacs, Simm, Ortner, and Cs{\'a}nyi]{batatia2022mace}
Batatia,~I.; Kovacs,~D.~P.; Simm,~G.; Ortner,~C.; Cs{\'a}nyi,~G. MACE: Higher order equivariant message passing neural networks for fast and accurate force fields. \emph{Advances in Neural Information Processing Systems} \textbf{2022}, \emph{35}, 11423--11436\relax
\mciteBstWouldAddEndPuncttrue
\mciteSetBstMidEndSepPunct{\mcitedefaultmidpunct}
{\mcitedefaultendpunct}{\mcitedefaultseppunct}\relax
\EndOfBibitem
\end{mcitethebibliography}

\end{document}